Shared and distinct exonic parts in alternative paths of splicing bubbles.

Daniel Witoslawski, Jelard Aquino, Chuanchuan He, Mira V. Han

Abstract
Detecting alternative splicing from complex splicing bubbles that contain more than two transcript paths are challenging. We present GrASE, a framework that distinguishes alternative path groups by the distinct and shared exonic parts they contain. Using lower-set bipartitioning, which efficiently enumerates downward-closed sets of paths, we find bipartitions of complex bubbles with well-defined distinct exonic parts. Applied to the human genome, GrASE tests 496,565 path groups from valid bipartitions across 22,299 genes, including alternative TSS and TTS events that are inaccessible to junction-based methods. GrASE can trade recall for higher precision and still detect substantially more events, due to its expanded test universe.

# Introduction

Alternative splicing (AS) enables different RNA isoforms to be generated from the same gene, expanding the functional capacity of a finite genome. AS can affect the mRNA stability, localization or translation, or can lead to different protein isoforms with diverse functions. Computational methods have been developed to detect AS from RNA-seq data. The splice graph, a Directed Acyclic Graph (DAG) that represents the splicing patterns of a gene was important in the early formulation of the problem and allowed the development of efficient solutions (Heber *et al.* 2002, Xing, Resch, and Lee 2004). In this framework, the complete set of transcripts can be represented by the possible paths along the DAG, and local AS events can be defined by the bubbles in the graph that represent alternative paths between two nodes. Nested bubbles can lead to structures with multiple alternative paths that quickly increase in complexity. So, the initial methods focused on simple localized events that have only two alternative paths that are easily classified and quantified. For example, rMATS(Shen *et al.* 2014), MISO (Katz *et al.* 2010) or SUPPA2(Trincado *et al.* 2018), while widely used, are restricted to specific binary events, such as skipped exons (SE), alternative 5'/3' splice sites (A5SS, A3SS), mutually exclusive exons (MXE) and retained introns (RI), but do not support multi-path or complex events.

Several tools have since been developed to overcome these limitations and expand the detection to complex events. Tools like JUM(Wang and Rio 2018) or MAJIQ(Vaquero-Garcia *et al.* 2016) approached the problem by focusing on each donor or acceptor independently, instead of looking at the whole bubble. *AS sub-structures* in JUM or *local splicing variations* in MAJIQ are similar concepts that are defined as sets of splice junctions sharing a common donor or a common acceptor. This allowed them to detect and quantify complex, multi-junction events, but they gave up on resolving the whole paths from the donor to the acceptor. Whippet(Sterne-Weiler *et al.* 2018) approached the problem by focusing on the local bubbles surrounding target nodes (exonic parts) that are limited by a span – defined by the farthest upstream and downstream nodes that are directly connected to the target node. This approach elegantly defines a localized complex AS event, but the detection relies on quantifying individual sub-transcript paths that cannot be directly observed, so Expectation-Maximization is used for this purpose. The common strategy among these solutions is to include complex events with more than two alternative paths, but limit the boundaries of the local subgraphs to prevent an explosion in complexity.

Here, we propose a generalized approach to complex splicing events that attempts to globally examine all bubbles in the graph. We sort all bubbles in the graph in partial order based on hierarchy, and examine them up to a threshold size, while optionally collapsing the internal bubbles as we move up the hierarchy. Then, to generate the comparisons among the $k$ possible paths (the alternatives in alternative splicing), we consider three different approaches. i) all the $C(k, 2)$ pairwise comparisons of $k$ paths ii) each path as a multinomial outcome iii) valid bipartition of $k$ paths. We find AS events by identifying shared and distinct parts in each of these comparisons, quantifying these parts, and comparing them across samples. We show that although we are forced to skip a small proportion of larger bubbles in the graphs, the approach examines more AS bubbles compared to existing methods. Detection relies on the quantification of shared and distinct exonic parts instead of sub-transcript paths, which obviates the need for estimation and uses direct observation of mapped reads.

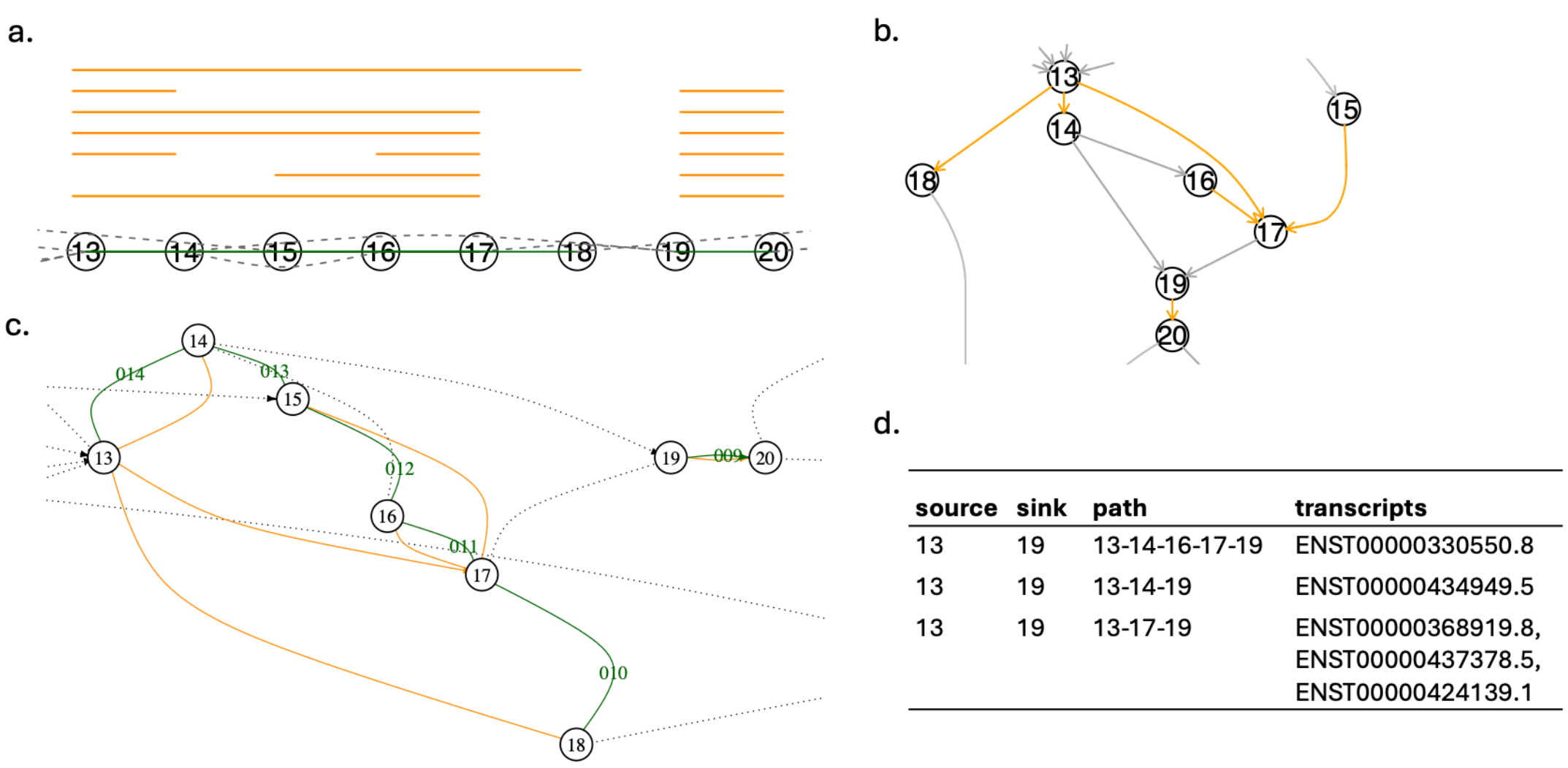

| source | sink | path | transcripts |
|---|---|---|---|
| 13 | 19 | 13-14-16-17-19 | ENST00000330550.8 |
| 13 | 19 | 13-14-19 | ENST00000434949.5 |
| 13 | 19 | 13-17-19 | ENST00000368919.8, ENST00000437378.5, ENST00000424139.1 |

***Figure 1. Partial transcript structure and its corresponding splicing subgraph.*** *a) Transcript isoform structure of gene SLC22A16 (ENSG00000004809) spanning nodes 13 to 20 (genomic positions 110,457,017-110,446,873). b) Traditional splicing graph showing only exon and intron edges. c) GrASE graph with exons, introns and additional exonic part edges that subdivide each exon. d) A complex bubble defined between nodes 13 and 19 with three alternative paths and the transcripts that traverse the respective paths.*

### Structuring Isoform Comparisons

We introduced a splicing graph, that we called GrASE (Aquino *et al.* 2025), that follows the convention as implemented in the SplicingGraphs package in Bioconductor (Bindreither *et al.*, n.d.), but has been extended to include the exonic part edges along the genome. GrASE is a directed acyclic graph (DAG) where vertices (nodes) represent the splicing sites for a given gene, ordered by their position from 5' to 3'. There are three types of edges, exons, introns, and exonic parts that connect the splicing sites, with orientations following the 5' to 3' direction. Our splicing graphs are constructed for each gene separately, based on reference gene annotations from Gencode (ver 34).

After constructing the splicing graph, we identify alternative splicing as the complete set of bubbles based on the gene annotations. Valid bubbles are structures in the graph where there exist at least two distinct valid paths – a path followed by at least 1 annotated transcript – between two nodes of a splicing graph. The nodes at the ends of the bubble are called "source" and "sink", and the number of distinct valid paths is the "size" of the bubble. Bubbles can be nested or can be overlapping without being nested.

Once we identify the set of AS events as bubbles, we need to decide how to structure the comparison among the multiple alternative paths so we can compare the read coverage. JUM (Wang and Rio 2018) or MAJIQ (Vaquero-Garcia *et al.* 2016) used a multinomial strategy where they quantify and compare one out of multiple outcomes. Whippet's approach focusing on a target node (exonic part) naturally leads to an intuitive partitioning of multiple paths into two clusters—inclusion paths vs. exclusion paths based on the inclusion of the target node (Sterne-Weiler *et al.* 2018). In this paper, we propose three different ways to look at the problem.

**(i) Pairwise $C(K,2)$ comparisons** - We compare every pair of individual paths with transcript annotation, generating $C(K,2) = \frac{K(K-1)}{2}$ pairwise comparisons. This approach provides the most granular analysis, revealing all pairwise differences between individual isoforms. However, it scales quadratically with the number of paths.
**(ii) multinomial comparisons** - We organize the $K$ paths into $K$ singleton groups, treating each path as a distinct outcome of the alternative splicing event. This yields exactly $K$ outputs (one per path), which we evaluate jointly in a single multinomial analysis, making it more efficient than enumerating all pairs. A limitation is that this one-vs-rest strategy will yield many paths lacking path-unique exonic parts, as we describe in the results. If path-unique parts are missing for even a single path, the multinomial comparison as defined here is not applicable**.**
**(iii) bipartition (two-group split) comparisons** - We partition the $K$ paths into exactly two groups yielding differential exonic parts by enumerating the valid bipartitions of the multiple alternative paths. Note that the number of nontrivial bipartitions grows exponentially as $2^{K-1} - 1$, so constraints on validity are helpful for tractability. We describe what valid bipartitions mean in more detail below.

**Lower set bipartitioning of paths**
With complex bubbles that have $K$ paths, there are potentially $2^{K-1} - 1$, ways to partition the alternative paths into binary classes. One may think that one intuitive way to partition the paths is a one-vs-rest (OVR) strategy, where each path is split from the rest into two partitions. This strategy is frequently used to transform a multinomial (multi class) problem to a bipartition (binary class) problem in biological contexts, *e.g.* Is this a T-cell or not? Does this species exist in this metagenome or not? Unfortunately, OVR partitioning is not always meaningful in the case of partitioning alternative splicing paths due to their nested relationships. The purpose of the path partitioning here is to identify exonic parts that are distinct between the two group of alternative a.

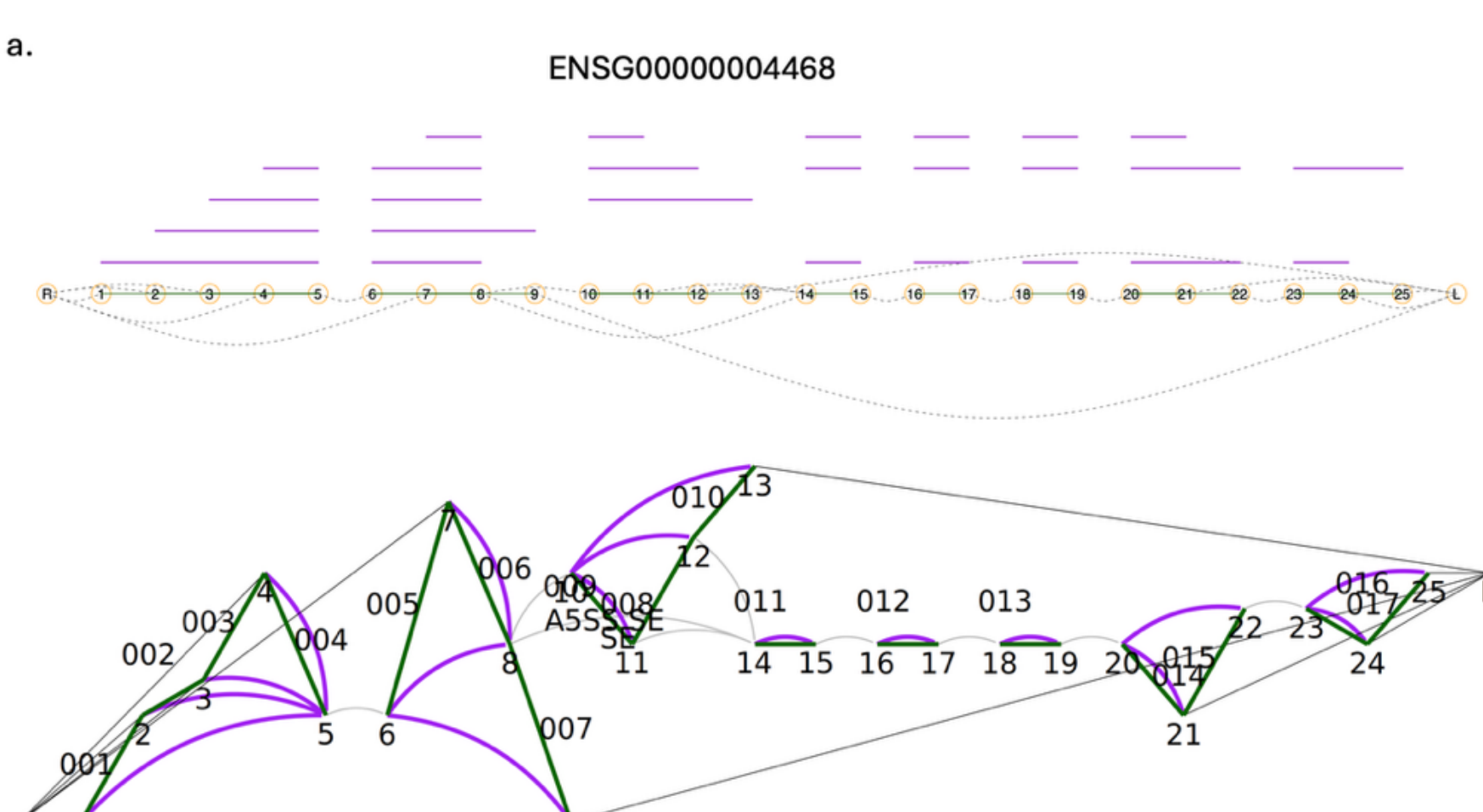

b. **bubble R-5**

**Alternative paths constituting the bubble R-5**

| vertex path | exon path | exonic part path |
|---|---|---|
| R-1-5 | $intron_{R,1}$-$exon_{1,5}$ | $intron_{R,1}$-$ex_part_{001}$-$ex_part_{002}$-$ex_part_{003}$-$ex_part_{004}$ |
| R-2-5 | $intron_{R,2}$-$exon_{2,5}$ | $intron_{R,2}$-$ex_part_{002}$-$ex_part_{003}$-$ex_part_{004}$ |
| R-3-5 | $intron_{R,3}$-$exon_{3,5}$ | $intron_{R,3}$-$ex_part_{003}$-$ex_part_{004}$ |
| R-4-5 | $intron_{R,4}$-$exon_{4,5}$ | $intron_{R,4}$-$ex_part_{004}$ |

c.
**Exonic parts in one-vs-rest bipartitioning of paths**

| Partition 1 | Partition 2 | distinct exonic part | shared exonic part |
|---|---|---|---|
| {001,002,003,004} | {002,003,004},{003,004},{004} | 001 | 004 |
| {002,003,004} | {001,002,003,004},{003,004},{004} | NA | 004 |
| {003,004} | {001,002,003,004},{002,003,004},{004} | NA | 004 |
| {004} | {001,002,003,004},{002,003,004},{003,004} | 003 | 004 |

d.
**Exonic parts in lower set bipartitioning of paths**

| Partition 1 | Partition 2 | distinct exonic part |
|---|---|---|
| {001,002,003,004} | {002,003,004},{003,004},{004} | 001 |
| {001,002,003,004},{002,003,004} | {003,004},{004} | 002 |
| {001,002,003,004},{002,003,004}, {003,004} | {004} | 003 |

***Figure 2 Lower set bipartitioning identifies distinct exonic parts where one-vs-rest bipartitioning fails.*** *a) The transcript isoform structure of gene ENSG00000004468 and its corresponding GrASE splicing graph. b) bubble R-5, a complex bubble with four alternative paths (R-1-5, R-2-5, R-3-5, R-4-5) corresponding to four nested transcript isoforms. The four exonic part paths form a strict containment chain: {004} ⊂ {003,004} ⊂ {002,003,004} ⊂ {001,002,003,004}. c) In one-vs-rest (OVR) bipartitioning, two of the four partitions yield no distinct exonic part (NA): when path {002,003,004} or {003,004} is placed alone in Partition 1, the opposing Partition 2 contains both its supersets and its subsets, so no exonic part is exclusively present in one partition and absent from the other. d) In lower set bipartitioning, only downward-closed subsets of the containment order are used as valid partitions, producing three partitions each with a well-defined distinct exonic part (001, 002, 003 respectively) that can be used downstream for quantifying differential exon usage.*

paths so we can compare their usage. The paths can be represented as sets of exonic part edges by dropping the intron edges which are irrelevant. The distinct exonic parts are edges that are present in all sets within one partition and entirely absent from the other partition. But, as seen in Figure 2, if one partition contains both the supersets and the subsets of the paths present in the other partition, as in certain OVR partitioning, then no such distinct exonic parts will be available.

To find meaningful partitions that lead to distinct exonic parts, we need to find partitions that respect the subset relationships among the paths: if a set is present in a partition, all the subsets are also present in the same partition. To solve this, we define an order among the paths in a bubble based on the subset relationship (*i.e.* subset to superset order). We then enumerate all proper downward-closed subsets (lower sets) (Davey and Priestley 2002) and return each as a valid binary partition: the lower set and its complement. Instead of exhaustively listing all powersets of the paths and checking its downward closedness, we used a more efficient algorithm described below. The algorithm systematically generates all the lower sets by recursion over the paths in subset-to-superset (topological) order along the containment graph, ensuring that a path is included only once all its subsets are present.

```
Input:  dag = containment graph with paths as nodes, edges connecting subset → superset
Output: list of valid partitions (downward-closed subsets)
---
nodes = all path names in dag
n     = number of nodes (paths)
topo = topological sort of dag (subsets come before supersets)

For each node v in topo:
  preds[v] = list of direct predecessors of v (nodes with edges into v)

Initialize empty list valid_splits
Define recursive function recurse(i, current_set):
  If i > n:
    If current_set is non-empty and not full:
      Add indices of current_set to valid_splits
    Return

  v = topo[i]

  # Option 1: include v – only if all preds[v] are already in current_set
  If all preds[v] are in current_set:
    recurse(i + 1, current_set ∪ {v})

  # Option 2: skip v – unconditional (always runs, independent of Option 1)
  recurse(i + 1, current_set)

# Initial call
recurse(1, ∅)

Return valid_splits
```

*Figure 3 Algorithm for enumerating valid lower-set bipartitions. Pseudocode for the recursive algorithm that generates all downward-closed subsets (lower sets) of the containment graph over alternative paths in a bubble. Input is the containment DAG; output is the list of valid binary partitions.*

We also give users the option to collapse the internal bubbles in hierarchical order. We order the bubbles so that shorter and deeper (inner nested) bubbles come first, to examine the complete set of bubbles in the splicing graphs for each gene. As longer and outer bubbles are examined, the inner and shorter bubbles nested within will be examined redundantly as part of larger bubbles many times. To avoid this redundancy, we can collapse the bubble as we traverse the bubbles from inner to outer order. After examining the inner bubble to identify the distinct and shared exonic parts, we collapse the inner bubble by retaining only the single path followed by the largest number of transcripts based on annotation, before moving to the outer bubble. The default is not to collapse the inner bubbles, and the results below are described without collapsing.

**Shared and distinct components in comparison of alternative splicing paths.**
In the simple binary AS events defined as A3SS, A5SS, SE, or RI, there are inclusion and exclusion reads that are counted separately. We can find analogous concepts in the exonic parts of complex events by identifying the exonic parts that are distinct between the comparisons, and exonic parts that are shared between the comparisons. This definition applies to all the comparison structures described in section "Structuring Isoform Comparisons".

To identify the distinct and shared exonic parts between comparisons, we follow the logic below: $partition1 = \{P_1, P_2, \dots\ P_m\}$, $partition2 = \{Q_1, Q_2, \dots\ Q_n\}$ , where $P_i$ and $Q_j$ represent sets of exonic part edges for each path in the bubble.

$I_1 = \bigcap_{i=1}^{m} P_i,\ U_1 = \bigcup_{i=1}^{m} P_i$
$I_2 = \bigcap_{j=1}^{n} Q_i, U_2 = \bigcup_{j=1}^{n} Q_i$
$I$ and $U$ are the intersection and the union of paths in each partition.

$$D_1 = \bigcap_{i=1}^{m} P_i - U_2$$
$$D_2 = \bigcap_{j=1}^{n} Q_j - U_1$$

$D_1$ and $D_2$ captures the distinct exonic parts that are exclusively shared by all paths in one partition but not the other partition.

$$S = I_1 \cap I_2 = \bigcap_{i=1}^{m} P_i \cap \bigcap_{j=1}^{n} Q_i$$

$S$ captures the shared exonic parts that are universally common in all paths of both partitions**.** In addition to the exonic parts shared universally within the bubble, we also find the incident edges incoming to the source, and outgoing from the sink, that are shared by all transcripts going through the bubble, to find additional shared exonic parts directly connected to the bubble.
Although, the logic has been described for two partitions, it is generally applicable for the case of pairwise $C(k, 2)$ comparison and multinomial comparison. For pairwise comparison, for each pair $(i, j)$, $D_1$ reduces to $D_i^{(i,j)} = p_i \setminus p_j$ and $D_2$ to $D_j^{(j,i)} = p_j \setminus p_i$. For multinomial comparison, for each path $k$, $D_1$ reduces to $D_k = p_k \setminus \bigcup_{j \neq k} p_j$, which captures exonic parts present in path $k$ but absent from all other paths.

**Read count representation for bipartition comparison**
Reads mapping to each exonic edge are quantified with HTSeq (Anders, Pyl, and Huber 2015) as part of the DEXSeq (Anders, Reyes, and Huber 2012) pipeline. Read counts mapping to distinct sets $D_1$ and $D_2$ and shared set $S$ are quantified by summing the reads that map to the exonic edges that are part of each set. Each non-empty distinct set ($D_1$ or $D_2$), paired with the shared set $S$, defines one path group, so one bipartition contributes up to two tests of path proportion. As we will see in the results, for many bubbles one of the distinct sets is an empty set, where the path skips the whole bubble and directly connects source to sink with an intron edge. If $D_1$ or $D_2$ is $\emptyset$, we omit the test for that path group. The total number of bipartition path proportion tests for a gene will be equal to the number of non-empty (D, S) path groups across the valid bipartitions of all of its bubbles.
For sample $i$ and path proportion test $j$, let $y_{D_{ij}}$ and $y_{S_{ij}}$ be the reads from sample $i$ mapping to the distinct set $D$ and the shared set $S$ for test $j$, respectively. The reads on the distinct set are treated as successes and the reads mapping to both the distinct and shared sets as trials.
$y_{ij} = y_{D_{ij}},\ n_{ij} = y_{D_{ij}} + y_{S_{ij}}$
The per-sample ratio $\pi_{ij} = \frac{y_{D_{ij}}}{y_{D_{ij}} + y_{S_{ij}}}$ is the $Path\ Proportion(\pi)$, analogous to the Percent-Spliced-In of junction-based methods. Note that the two test arising from the same bipartition share the same $y_{S_{ij}}$, but each is fitted and reported separately.

**Statistical models for bipartition comparison: beta-binomial model**
For each sample $i$ and partition $j$, GrASE models the read counts $(y_{ij}, n_{ij} - y_{ij})$ using a beta-binomial distribution implemented in glmmTMB (Brooks *et al.* 2017).

$$y_{ij} \sim BetaBinomial\ (n_{ij}, \pi_{ij}, \phi_j)$$

Where $\pi_{ij}$ is the *path proportion* $\frac{y_{D_{ij}}}{y_{D_{ij}} + y_{S_{ij}}}$ and $\phi_j$ is the test specific overdispersion parameter. glmmTMB utilizes the parameterization where the overdispersion parameter $\phi$ is related to the beta-binomial correlation parameter $\rho$ by $\rho = 1/(1 + \phi)$, and where larger $\phi$ corresponds to lower overdispersion. A likelihood ratio test (LRT) comparing the full model $logit(\pi_{ij}) \sim \mu_j + \beta_j x_i$ ,where $x_i$ indicates the sample grouping, against the null model $logit(\pi_{ij}) \sim \mu_j$, is used to detect differential exon usage between sample groups.

**Empirical Bayes moderation of dispersion**
To obtain first pass per-test dispersion estimates, an intercept-only beta-binomial model is fitted to each test independently using glmmTMB. The dispersion $\widehat{\phi_j}$ is extracted via function *sigma* and its variance on the log scale is obtained by the delta method from the full variance–covariance matrix of the fitted model:

$$\widehat{Var}\left(log\widehat{\phi_j}\right) = diag(\Sigma)_{disp}$$

and

$$\widehat{Var}\left(\widehat{\phi_j}\right) = \widehat{\phi_j}^2 \cdot \widehat{Var}(log\widehat{\phi_j})$$

Raw per-test dispersion estimates ($z_j = log\widehat{\phi_j}$) are noisy, especially for tests with low read depth or few samples. GrASE borrows strength across tests by shrinking log-transformed dispersion estimates as originally described in Smyth (Smyth 2004). Optionally, a trend-based prior can be used in place of the global mean similar to DESeq2 (Love MI, Huber, and Anders 2014). A loess curve is fitted to $log\widehat{\phi_j} \sim log(\overline{n}_j)$ (log mean coverage per test), using the most reliable 90% of estimates by sampling variance. The trend prediction $\hat{z}_{trend}(j)$ replaces $\bar{z}$ as the shrinkage target, so that the prior adapts to the mean-dispersion relationship in the data.

We implement two Empirical Bayes methods for computing the moderated $\tilde{z}_j$ : *EBapprox*, a closed-form shrinkage estimator based on a normal-normal empirical Bayes shrinkage on log-dispersion (Smyth 2004), and *EBmap*, a MAP estimator based on the actual beta-binomial likelihood optimization (Supplementary Materials). An *MLE* model without any EB shrinkage is also implemented to serve as a reference baseline comparison. For both *EBapprox* and *EBmap*, the moderated log-dispersion $\tilde{z}_j$ is used to fix $\phi_j$ at a common value in both the null and full beta-binomial models, implemented via the glmmTMB parameter mapping mechanism. The LRT statistic $\Lambda = 2(l_1 - l_0)$ with $\phi_j$ fixed is compared to a $\chi^2$ distribution ($d.f.$ =1) to obtain a *p*-value.

**Statistical models for Multinomial comparisons**

For multinomial events, per-sample counts are represented as a vector across all distinct paths, which is modelled jointly. A Dirichlet-multinomial (DM) model is used to model the vector of multiple alternative paths in a bubble.

**Precision estimation.** For each multinomial event $j$, let $y_{ijk}$ denote the reads from sample $i$ mapping to path category $k$, where $k = 1 \dots K$, and $K$ is the number of paths. "event" here means the whole multinomial comparison, so there is one event for one bubble. DM precision $\alpha$ is estimated per event via direct one-dimensional log-likelihood optimization over $log(\alpha_j)$. Under the intercept-only null model the MLE of the proportion vector is the empirical marginal $\hat{\pi}_{jk} = \sum_i y_{ijk} / \sum_{ik} y_{ijk}$, so precision estimation reduces to a scalar search. The variance of $log(\hat{\alpha}_j)$ is obtained from the analytical observed Fisher information (the negative reciprocal of the second derivative of the log-likelihood with respect to $log(\alpha_j)$ at the MLE). Precision estimates are moderated by the same the Normal-Normal Empirical Bayes shrinkage (Supplementary Materials), working in log-precision space.

**Direct DM LRT with fixed EB precision.** With precision fixed at the EB-moderated value $\alpha_j = exp(\tilde{z}_j)$, both null and alternative log-likelihoods are evaluated directly in closed form and no iterative fitting is required. Under the null model the common proportion MLE is $\hat{\pi}_{jk} = \sum_i y_{ijk} / \sum_{ik} y_{ijk}$. Under the alternative, per-group proportion MLEs are the group-specific empirical marginals $\hat{\pi}_{jk}(g) = \sum_{i \in g} y_{ijk} / \sum_{i \in g, k} y_{ijk}$. The LRT statistic $\Lambda = 2(l_1 - l_0)$ is compared to a $\chi^2$ distribution with $(K-1)(G-1)$ degrees of freedom, where $K$ is the number of path categories and $G$ is the number of groups.

**Other statistical considerations**

**Non-parametric fallback.** A Wilcoxon rank-sum test on per-sample proportions ($y_i/n_i$) is available as a model-free alternative. The effect size is reported as the difference in group means of the proportion.

**P-value combination for bipartition tests.** Each bipartition bubble is tested from both sides ($D_1$ *vs.* $S$ and $D_2$ *vs.* $S$), corresponding to the two possible distinct paths. Two combination strategies are optionally applied: (i) the minimum of the two *p-values* is taken as the bipartition-level statistic (min-p combination), and (ii) Fisher's combined probability test is applied ($X = -2 \sum_k p_k \sim \chi^2_{2k}$).

**Independent filtering**. Before the nested BH described below, we use a DESeq2-style independent filtering (Love MI, Huber, and Anders 2014) based on the BaseMean, which is the mean of total read counts ($y_{D_{ij}} + y_{S_{ij}}$). Tests whose baseMean falls below the data-driven threshold are masked by setting their p-value to NA, excluding them from the multiple-testing pool.

**Multiple testing correction.** GrASE uses nested Benjamini–Hochberg (BH) that screens genes, then localizes path proportion tests within the genes that pass (Van den Berge *et al.* 2017). In the first stage that screens genes, the *K* test p-values of a gene are combined into one gene-level p-value by Simes' method, $p_{\text{Simes}} = \min_j (K * p_j / j)$ (Simes 1986). BH is then applied across the one-per-gene Simes p-values at alpha = 0.05, giving the column *padj_gene*. This controls the FDR of the set of genes with at least one differential path proportion. In the second stage that localizes path groups, only for genes with *padj_gene* < 0.05, plain BH is applied to that gene's path proportion test p-values, giving the test-level column *padj*. Path proportion tests in genes that fail Stage 1 get their *padj* set to NA.

**Simulation-based benchmarking of GrASE statistical models**

To evaluate the performance of GrASE statistical models for detecting differential alternative splicing (AS), we conducted a simulation study based on the simulated data from Love et al. (Love M, Soneson, and Patro 2018) in which known ground-truth differential events were introduced into RNA-seq data. To briefly describe the simulation design, genes were classified into four simulation types: genes with no induced change (Background; n = 53,461), genes with differential gene expression only (DGE; n = 1,501), genes with differential transcript expression (DTE; n = 1,501), and genes with differential transcript usage (DTU; n = 1,501).

The simulation was designed to evaluate the detection at the transcript level, which is different from our goal of detection at the exonic part level. To translate from transcript to exonic part, we assigned exonic parts to positive or negative calls based on their transcript membership. For DTE, one transcript has its expression scaled by a fold change of 2~6x in one condition. All exonic parts overlapping the DTE transcript were labeled positive. For DTU, two transcripts swap their expression levels between conditions, with no net change in total gene expression. An exonic part is labeled positive for DTU only if one of the swapped transcripts overlap it and the other counterpart transcript does not, meaning the exonic part distinguishes the two transcripts. Exonic parts shared by both swapped transcripts or by neither are labeled negative. This stricter criterion means DTU produces fewer positive exonic parts per gene than DTE. Note that the DTE assignment is structurally biased against our approach since all exonic parts that are part of the transcript is considered positive, but we only test exonic parts that are distinct parts of bubbles.

To translate from transcript to splice junctions for rMATS, we assigned junctions as positive or negative based on the structural overlap with junction-based transcript matching derived from the fromGTF.*.txt (full annotated universe, ~61k events). DTE changes a single transcript's abundance, so it is Positive in all junctions where that transcript passes through either side (inclusion or exclusion). DTU swaps a pair in opposite directions, so it is positive only if the two swapped transcripts are not on the same side of inclusion or exclusion.

Because both GrASE and rMATS discern which events to test beforehand, testing against the full universe is unfair to these software. To assess this effect, we evaluate the performance based on two universes. For GrASE, (i) the full gene/exonic part set (57,964 genes, 615,358 exonic parts), in which all exons in null genes count as true negatives, and all exons in positive DTE transcripts count as true positive; (ii) a restricted universe comprising only exons that each tool explicitly tested, so that recall does not include structural un-testability. For rMATS, (i) the full annotated fromGTF events (14,795 genes, 60,044 events), including untested events. (ii) the restricted universe of events that are tested by rMATS, which exclude events with < 10 counts coverage per recommendation.

Performance was quantified using micro-averaged precision, recall, and F1 score computed over all exonic part calls, at adjusted p-value (padj) thresholds of 0.01, 0.05, 0.10, and 0.20. Unless otherwise noted, results are reported at padj <= 0.01. Throughout, TP, FP, FN, and the derived precision, recall, and F1 are counts of exonic parts, while "Universe (tests)" is the number of tests. The two units differ because a single test can yield multiple exonic parts and tests within a gene can share them. Counting true and false calls at the exonic parts level makes methods with different numbers of total tests directly comparable.

# Results

### Application to the human genome

We identified *N* = 263,308 bubbles across the 22,299 genes in the genome without collapsing the bubbles. The mean number of bubbles per gene was 12.8, and the median was 9. The span of a bubble is defined as the difference in topological rank between source and sink, or equivalently, the number of nodes from source to sink, not counting the source, in the 5'-to-3' direction. The median span of a bubble was 14, and the median number of paths per bubble was $k = 4$. To maintain computational efficiency, we limited the size of the bubbles to a maximum of $k = 15$ alternative paths. This excluded 7,985 of the most complex bubbles (~3 % or the total *N*=263,308 bubbles).
We can distinguish between internal bubbles, which correspond to the standard Alternative Splicing (AS) events, and peripheral bubbles that involve the R (5' end) or L (3' end) nodes, which correspond to alternative Transcription Start Site (TSS) or alternative Transcription Termination Site (TTS) events. Software that relies on reads spanning both junctions cannot detect TSS or TTS events as these sites are either starts that have no upstream incoming junction or stops that have no downstream outgoing junction. In contrast, since GrASE utilizes coverage across exonic parts, it processes AS and alternative TSS/TTS events identically, enabling the detection of both. As reported in the literature, bubbles defining TSS/TTS events (192,199) significantly outnumber bubbles defining internal AS events (71,109) in our data. Thus, this detection capability represents a unique strength of the GrASE framework.

Applying the bi-partitioning algorithm above, 244,337 out of 263,308 bubbles yielded at least one valid bipartition, resulting in a total of 1,066,383 valid partitions. For nested bubbles that yield the same $D_1$ and $D_2$, we removed the duplicated bipartitions and keep only the row with smallest $|S|$ per unique ($D_1$, $D_2$). If $|S|$ is the same, we prioritized the $S$ with the exons closest in genomic order to the exons in $D_1$and $D_2$. Deduplicating these bipartitions that yield the same ($D_1$, $D_2$) reduced the total number of lower set partitions to 496,565 with defined distinct and shared exonic parts. Of those, there were about fifteen times more partitions defined for TSS/TTS events (464,614) than partitions defined for internal AS events (31,951). Evaluating all possible pairwise path combinations $C(K, 2)$ yielded a total of 3,376,146 valid pairwise comparisons. Deduplicating these pairwise comparisons with the same logic as above—using the same criteria for $|S|$ and genomic proximity—reduced the total to 1,682,159 comparisons with defined distinct and shared exonic parts. Of those, there were about twelve times more comparisons defined for TSS/TTS events (1,554,376) than for internal AS events (127,783). Applying the multinomial algorithm, we found 247,613 total multinomial comparisons. But, filtering out comparisons where at least one path has no distinct exonic part and deduplicating the comparisons reduced the total to 50,859 comparisons. Comparing all paths in a bubble simultaneously in multinomial comparison, it is more likely that at least one path has no exonic parts distinct from all other paths combined, causing the bubble to be filtered out. Due to this, the number of bubbles that could be tested with multinomial were significantly smaller than other comparisons. Unlike other comparisons, there were about 1.5 times more comparisons defined for internal AS events (30,155) than for TSS/TTS events (20,704).

### Reduction in complexity

The algorithm based on the containment graph (Figure 3) showed significantly reduced computational complexity compared to exhaustively checking all the powersets of the paths which has exponential cost. With $K$ nodes (paths) and $m$ edges (subset relation) in the containment graph defined per bubble, the complexity of the topological sort is linear to the size of the graph $O(K+m)$, and the complexity of the recursive function is proportional to the sum of the sizes of valid lower sets. The overall complexity is $O((K+m)*L(b))$, where $L(b)$ is the number of lower sets for a given bubble $b$.
In most splicing graphs with nested structures, we found that the total number of lower sets are significantly smaller than the naive enumeration of the powerset ($2^{K-1}$-1) required for examining every possible bipartition. Table 1 shows the number of lower sets observed for the human genome annotation (gencode v28).

***Table 1 Lower set sizes for splicing graph bubbles by number of paths***. *For each bubble size (K), we report the number of bubbles with the given size identified in the human genome (N), and the summary statistics of the lower set size L(b) for those bubbles. The naive enumeration size based on the powerset $2^{k-1}$-1 are shown alongside L(b) for comparison. ratio = powerset / mean L(b).*

| *K* (paths) | *N* (bubbles) | mean *L(b)* | median *L(b)* | max *L(b)* | powerset | ratio |
|---:|---:|---:|---:|---:|---:|---:|
| 2 | 83247 | 1 | 1 | 1 | 1 | 1.00 |
| 3 | 41187 | 2.23 | 2 | 3 | 3 | 1.35 |
| 4 | 29456 | 3.61 | 3 | 7 | 7 | 1.94 |
| 5 | 23132 | 5.07 | 5 | 11 | 15 | 2.96 |
| 6 | 17848 | 6.56 | 6 | 16 | 31 | 4.73 |
| 7 | 13418 | 8 | 8 | 19 | 63 | 7.88 |
| 8 | 10151 | 9.45 | 9 | 20 | 127 | 13.44 |
| 9 | 7427 | 10.86 | 11 | 22 | 255 | 23.48 |
| 10 | 5598 | 12.14 | 12 | 22 | 511 | 42.09 |
| 11 | 4132 | 13.52 | 13 | 24 | 1023 | 75.67 |
| 12 | 3116 | 14.89 | 15 | 26 | 2047 | 137.47 |
| 13 | 2459 | 16.13 | 16 | 27 | 4095 | 253.87 |
| 14 | 1789 | 17.36 | 17 | 30 | 8191 | 471.83 |
| 15 | 1377 | 18.67 | 19 | 31 | 16383 | 877.50 |

92,640 (38%) of 244,337 bubbles have $L(b) = 2^{K-1}-1$, indicating there is no reduction in complexity for those specific instances. These include bubbles with only two paths ($K$=2) or bubbles that have more than two paths ($K$>2) but the structure is a simple linear chain. However, for $K >= 5$, the number of lower sets $L(b)$ to examine for each bubble is consistently about 3x to 878x smaller than the naive powerset. While the naive powerset enumeration grows exponentially, $L(b)$ grows roughly linearly with $K$ (mean $L(b) \sim K$).

In practical runs, larger bubbles can still take a long time. It took about twelve hours to run the whole human genome with 20 cores. So, we allow the users to specify the limit on the size of the bubbles that are examined. The default is to only examine bubbles with less than 15 alternative paths, which amounts to 97% of all bubbles in the human genome. The time that it takes to generate the partitions and identify the shared and distinct exonic parts is significant, but it only has to be run once for a set of annotations (gtf). We have run it for the human genome gencode ver. 28 and ver. 34 and other frequently used genomes and annotations, and the data is shared at XXX.

**Bipartition comparison shows best precision on simulated data among the three comparison structures.**

With shared and distinct parts representing alternative path groups identified, we could directly observe and quantify the reads mapping to distinct exonic parts to compare the usage of the different alternative path groups across samples. We compared the detection of alternative splicing based on the path proportion of distinct and shared exonic parts across the simulations of Love et al. The simulation has four categories. In DTE genes, it picks one expressed transcript and applies a fold change of 2-6x to that single transcript in one condition. In DTU genes, it picks the highest expressed transcript and one lower-expressed transcript, then swaps their TPMs between conditions. There is no net change in total gene expression, just the proportion changes. In DGE genes, it increases all transcripts with same fold change in one condition. Since all transcripts scale by the same factor, the proportions within the gene stay constant, only total gene expression changes. The background genes are all the rest of the genes where no change happens. DGE and Background genes are true negatives for splicing detection.

First, we compared the performance across different comparison structures we examined. Bipartition showed the best overall performance among the three comparison structures with highest precision, well-controlled FPs, and best DTU F1. DTE recall for bipartitions is low but this appears to be a general limitation across all three comparisons. $C(K, 2)$ comparison shows higher recall but with lower precision. Multinomial comparison has limited gene coverage (~23% fewer genes tested), which deflates its recall in the full analysis. In the restricted analysis, where we only count among the exonic parts tested by each method, the recall of multinomial comparisons rises to 0.523 (Figure 4b).

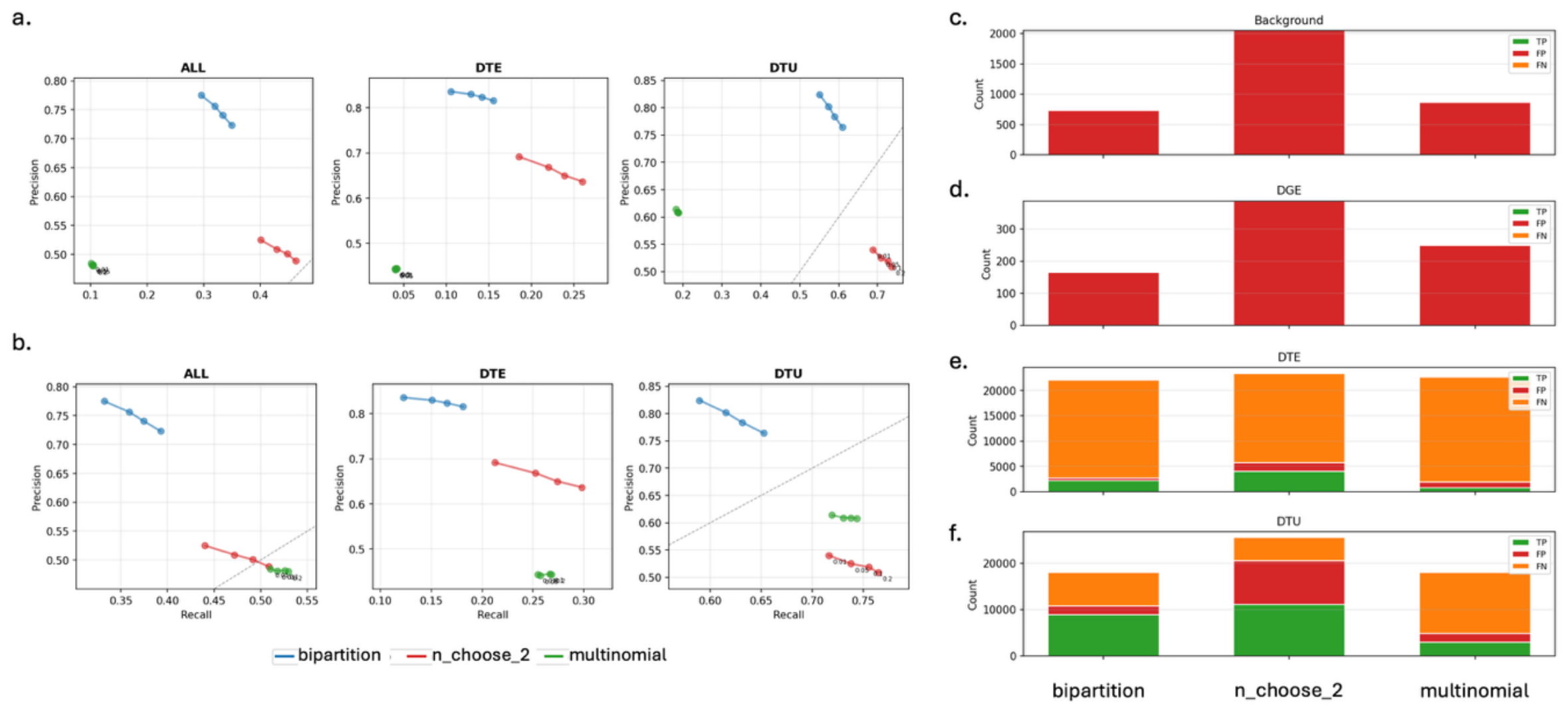

*Figure 4 a-b. Precision-recall curves for GrASE splicing event detection across three comparison structures, bipartition, $C(k, 2)$ and multinomial in a. total universe b. restricted universe of tested events. Each point represents a padj threshold (0.01, 0.05, 0.10, 0.20); Panels show performance on all events (ALL), differential transcript expression (DTE), and differential transcript usage (DTU). The diagonal in DTU indicates the precision = recall line. c-f. Confusion count stacked bar charts for three comparison structures at padj <= 0.01. Each bar shows the total number of true positives (TP, green), false positives (FP, red), and false negatives (FN, orange) per method across simulation categories (DTE, DTU, Background, DGE) in the unit of exonic parts.*

Control of False Positives (FP) is a key requirement for genome-wide use. We looked at the performance on events from null genes (Background and DGE), which carry no true difference in alternative splicing. At padj <= 0.01, The three comparison structures differed dramatically in their null-gene FP rates (Figure 4 c-d). Bipartition has the fewest FPs from Background and DGE (899 combined), indicating well-controlled type I error. $C(K,2)$ generates about 2.7x more false positives from Background and DGE (2,436) than bipartition and about 2.2x more than multinomial, reflecting its elevated false positive rate. multinomial has the largest False Negative counts (33,918) due to lower coverage. Based on these results, we conclude that bipartition is the better comparison structure to identify AS among complex bubbles. We report the rest of the results based on the bipartition comparison.

**Empirical Bayes shrinkage increases the precision compared to MLE or non-parametric test.**

The two Empirical Bayes (EB) bipartition models, EBmap and EBapprox, produce similar results across all evaluation metrics, with EBapprox slightly outperforming EBmap on precision. At padj < 0.01 the two models agree closely on false positive counts (Background: 734 vs 778; DGE: 165 vs 172). EBapprox has the higher DTU precision (0.824 vs 0.813) at comparable recall (0.550 vs 0.556) and F1 (0.660 vs 0.660). Because EBapprox gives equal-or-better precision at lower computational cost, we used EBapprox as the primary EB model for the rest of the analyses. The difference between the EB-based methods and MLE and Wilcoxon is substantial. MLE generates ~1.7x False Positives with 1,250 background FPs and 258 DGE FPs at padj < 0.01. This excess in null FPs is not compensated by better DTU performance as MLE achieves DTU precision of 0.816 and recall 0.520, both at or below the EB models. Without any kind of shrinkage, the MLE dispersion estimate is frequently collapsing to the anti-conservative binomial limit, inflating null false positives. Wilcoxon, by contrast, is the most conservative on null genes, as it produces the fewest Background FPs (561) and DGE FPs (144) of any model. But this comes at the cost of the lowest DTU precision (0.800) and recall (0.528), so its overall F1 (0.636) trails the EB models.

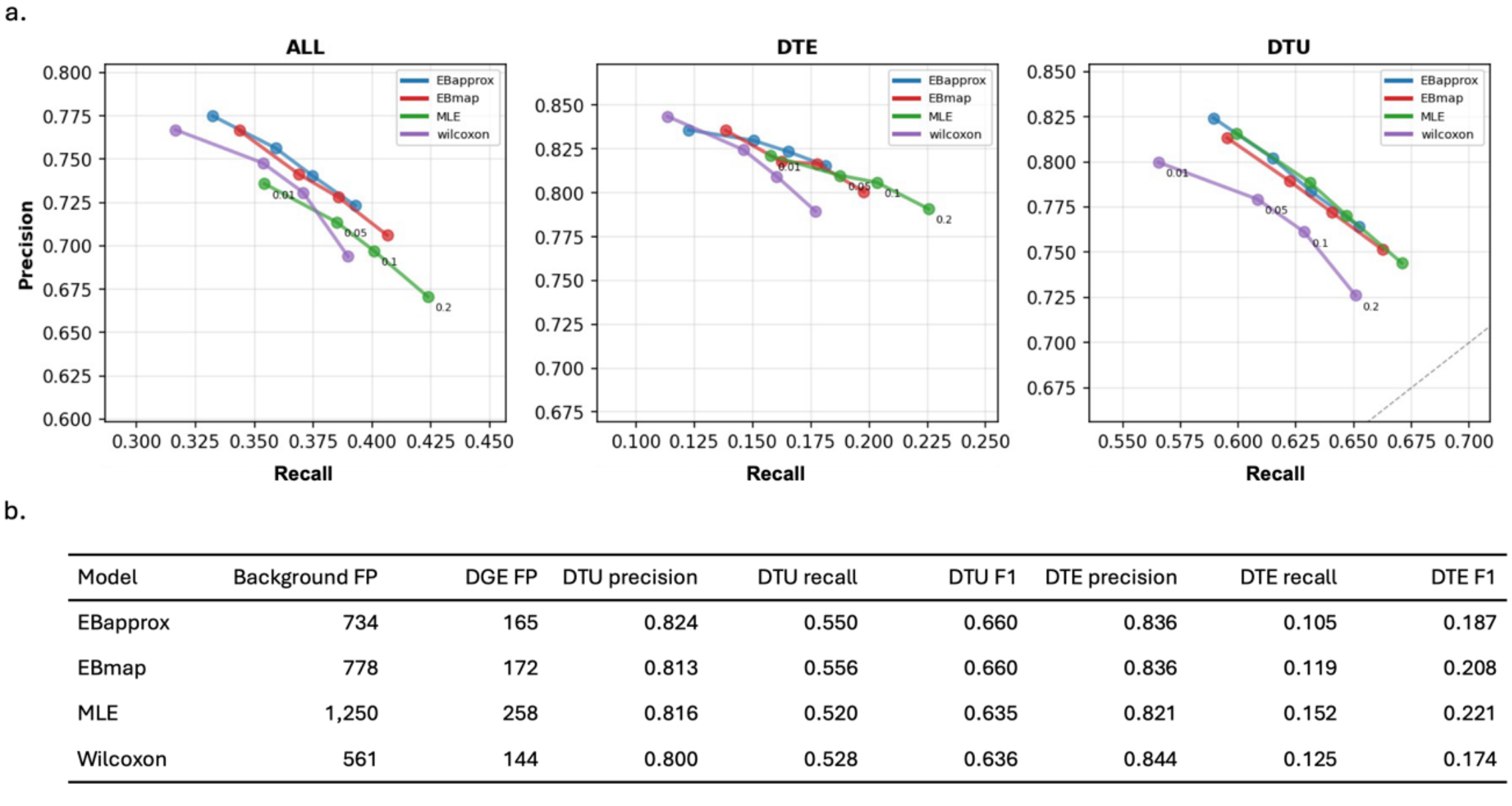

| Model | Background FP | DGE FP | DTU precision | DTU recall | DTU F1 | DTE precision | DTE recall | DTE F1 |
|---|---|---|---|---|---|---|---|---|
| EBapprox | 734 | 165 | 0.824 | 0.550 | 0.660 | 0.836 | 0.105 | 0.187 |
| EBmap | 778 | 172 | 0.813 | 0.556 | 0.660 | 0.836 | 0.119 | 0.208 |
| MLE | 1,250 | 258 | 0.816 | 0.520 | 0.635 | 0.821 | 0.152 | 0.221 |
| Wilcoxon | 561 | 144 | 0.800 | 0.528 | 0.636 | 0.844 | 0.125 | 0.174 |

*Figure 5. a. Performance comparison of statistical models used for differential usage between bipartitions of alternative paths. b. False positive counts and performance metrics at adjusted p-value = 0.01.*

On identifiable tests (where the data shows dispersion), EBapprox and EBmap recover essentially the same value of per-test dispersion, while on non-identifiable tests with no dispersion signal, each method supplies its own shrinkage-based estimate, so they diverge into a weak correlation (Supplementary Figure 1). EBapprox moderates more aggressively towards the common target than EBmap. The two EB models nonetheless agree closely on effect size (Supplementary Figure 2). Since the two tests for each bipartition shares the set S, we considered combing the two tests for each side of the bipartition, via Fisher's method or by the minimum p-value. Overall, combination methods offered a marginal DTU recall gain at a cost of substantially lower DTE precision and modestly higher null FPs (Supplementary Materials, Supplementary Figure 3). Since reporting each sides test separately provides the better precision-recall balance, we report each test separately.

**A post-hoc log2FC filter can correct false positives in DTE genes due to reference effect.**
In our initial analysis the False Positive rates on the DTE events were even larger than what is now reported in the main results. To understand the source of the FPs, we investigated the FP exons. In most FPs (72.3%), neither of the distinct exonic parts $D_1$ or $D_2$ contained exons of the changed transcript. Careful observation showed that these arise from a reference effect inherent to compositional testing. GrASE tests the path proportion $\pi_{ij} = \frac{y_{D_{ij}}}{y_{D_{ij}} + y_{S_{ij}}}$, for each path groups of the bipartition. We define the reference effect as a spurious shift in the detected $\pi_{ij}$ path proportion at unrelated bubbles due to the change in the shared exonic part count ($y_{S_{ij}}$) from transcript expression change. The isoform structure of genes have many nested bubbles that ultimately comprise the outermost bubble that starts at R and ends at L. Because of this nested structure, multiple bubbles can have the same exonic parts as the incident edge that is the shared exonic part or “reference” ($S$). When a gene contains a transcript that changes in absolute expression (DTE), the fold change affects every exonic part that it passes through. The transcript may pass through $S$ of multiple bubbles, without passing through the $D_1$ or $D_2$ of those bubbles. So, one DTE transcript can generate FPs across many bubbles simultaneously, which explains why in 72% of FPs, the changed transcript is in the $S$ of their bubbles, not in any of their $D_1$ or $D_2$. The path proportion $\pi_{ij} = \frac{y_{D_{ij}}}{y_{D_{ij}} + y_{S_{ij}}}$ shifts between conditions not because $y_{D_{ij}}$ changes, but because $y_{S_{ij}}$ changes. GrASE flags the bubble as significant, producing a False Positive. This effect is not specific to GrASE, and it applies equally to DEXSeq. The problem has been documented in the literature (DEXSeq Anders et al. 2012), and we have observed it before (GrASE original paper), but this is the first time the effect on FP is quantified at the exonic part level. We show a worked example in gene ENSG00000133706.17 (Figure 8). ENST00000504323.5 is a transcript covering E026-E038 in the middle of the gene ENSG00000133706.17 that is decreased by 5.7-fold as part of the DTE simulation. GrASE correctly identifies the exons of the changed transcript (E026-E038 region) as differential. However, two additional large bubbles covering the 5' region ($D$: E003-E023, 15 exons) and the 3' region ($D$: E039-E067, 25 exons) of the gene are also called significant. None of their $D_1$ or $D_2$ exons overlap with the changed transcript ENST00000504323.5. Instead, their shared exonic part $S$ (E029 and E030) overlaps with ENST00000504323.5. As the transcript decreases, $y_{S_{ij}}$ shrinks and $\pi_{ij}$ inflates. As the reference effect propagates to large bubbles that span multiple exonic parts on both sides, unrelated FPs outnumber TPs 7:1.

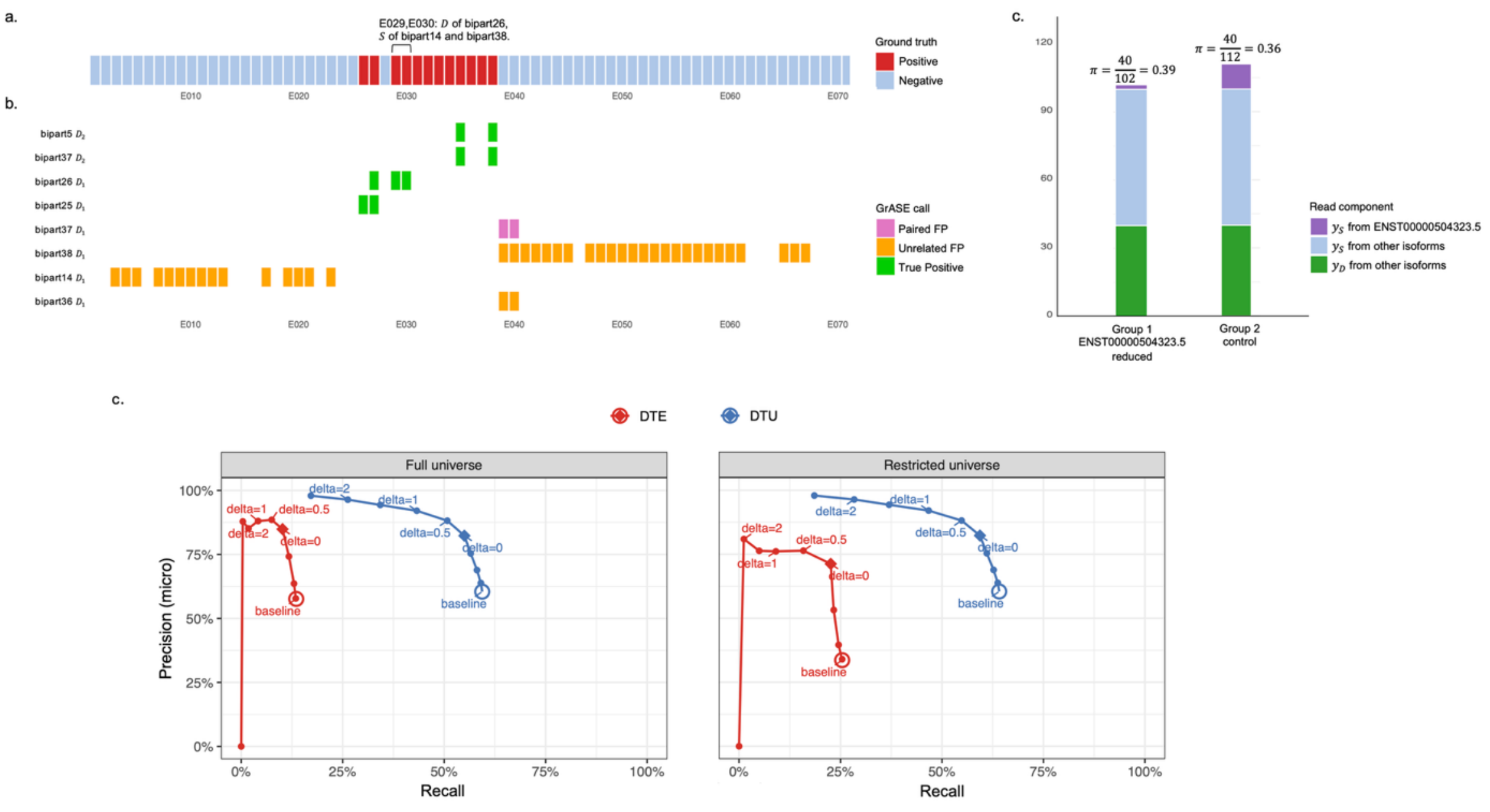

*Figure 6. False Positive arising from reference effect. a) Ground truth in exonic parts (E001-E071). As the transcript ENST00000504323.5 was reduced by 1/5.7x fold change as part of the DTE simulation, all exonic parts that are part of the transcript are considered Positive exonic parts (E026-E038) in the ground truth translation. The rest are negative exonic parts. b) Eight significant tests are called by GrASE at padj < 0.05. In True Positive calls distinct exons D overlap with the Positive ground truths in E026-E038 region. Paired FP are false positive calls where distinct exonic parts Dare from the other side partition of a TP bubble. Unrelated false positive calls are from bipartitions 38, 14, and 36, that share the shared exonic parts S. bipartition 38 $D_1$ covers the 3' region (E039-E067, 25 exons) and bipartition 14 $D_1$ covers the 5' body (E003-E023, 15 exons). c) Read counts from bipartition 38 $D_1$ ($D_1$ = E039-E067, S = E029-E030). Stacked bar chart showing read components for Group 1 (changed) vs Group 2 (control). The $y_D$ count is equal in both groups (~40 reads). The $y_S$ component from other isoforms passing S are also equal (~60 reads). The $y_S$ component from ENST00000504323.5 passing S shrinks from 11 reads (Group 2) to 2 reads (Group 1) because the transcript dropped ~5.7x. n decreases in Group 1, leading to $\pi$=40/102=0.39 while Group 2 $\pi$=40/111=0.36, so $\pi$ is higher in Group 1. GrASE calls a significant shift in path proportion, but no transcripts overlapping with $D_1$ (E039-E067) are changed in the simulation resulting in a False Positive. c. $lfc_{D_net}$ filter reduces False Positives.*

To address this, we propose a model-agnostic post-hoc filter that targets the problem directly. We computed for each comparison the log2 fold change of $y_{D_{ij}}$ ($lfc_D$) and log2 fold change of $y_{S_{ij}}$ ($lfc_S$) across conditions and reported the $lfc_{D_net} = abs(lfc_D) - abs(lfc_S)$. When $lfc_{D_net} \leq \delta$, $y_{S_{ij}}$ is changing more than $y_{D_{ij}}$, indicating the shift in path proportion $\pi_{ij}$ is likely driven by the shared reference rather than a genuine splicing change. We don't have to worry about normalization since library-size bias cancels in this comparison because $y_{D_{ij}}$ and $y_{S_{ij}}$ are measured from the same samples. The $lfc_{D_net}$ score is reported in the output and users can apply a $\delta$ threshold to filter calls.

Supplementary Table 1. lists all the comparisons for gene ENSG00000133706.17 before the filtering showing 44 FP exonic parts (4 tests) and 6 TP exonic parts (4 tests). As can be seen in column filtered_status, all 44 False Positives are filtered out by the post hoc $lfc_{D_net}$ filter. 3 out of 4 of the True Positive tests show positive $lfc_{D_net}$ (0.29-1.38) and are retained (TP), while 1 Positive test shows negative $lfc_{D_net}$ (-0.23) and fails to pass the filter (FN), resulting in 0 FPs and 3 TPs and 1 FNs after filtering.

Looking at the data genome-wide, the null FPs were significantly reduced with the post hoc filter on $lfc_{D_net}$ (Figure 6c). At padj < 0.01 and $\delta$ = 0, Background FPs fell from 1,057 to 734 (-31%) and DGE

FPs from 645 to 164 (-75%). DTE precision improved from 0.528 to 0.836, while DTE recall fell from 0.130 to 0.104 as some weaker true DTE signals also have $lfc_{D_net}$<= 0. For DTU, precision improved from 0.605 to 0.826, and DTU F1 improved from 0.596 to 0.660. However, recall dropped from 0.587 to 0.550, with 15.6% of DTU true positives getting removed by the filter, indicating that genuine DTU events can be incorrectly filtered out. We experimented with the $\delta$ threshold to examine the precision-recall tradeoff. At $\delta$ = 0.5, DTU precision reaches 0.879 at the cost of recall dropping to 0.507, and at $\delta$ = 1.0, precision reaches 0.917 but recall falls to 0.427. For most applications $\delta$ = 0 provided a reasonable balance.

### GrASE focuses on precision over recall

We also compared the performance of the best model in GrASE with existing software DEXSeq and rMATS. The simulation was controlled at the transcript level, by manipulating the fold change of specific transcripts, but detection is done at the exonic part level for GrASE and DEXSeq, or at the splice junction level for rMATS. We described how we establish the ground truth by translating transcript change to exonic part change or splice junction change in the methods section "Simulation-based benchmarking of GrASE statistical models". Based on the recommendations of rMATS documentation, events that fail the PSI filter (average inclusion level (PSI) per group within [0.05, 0.95]) have their FDR forced to 1 so they will contribute to the Negative calls but not to the Positive calls (TP or FP).

GrASE had highest overall precision (0.775 overall, 0.824 DTU) (Table 2). Its recall is comparable to DEXSeq, so GrASE's higher precision is not traded off with worse recall. DEXSeq has similar recall to GrASE but much lower precision (0.537), with far more FPs in background (1,729) and DGE (1,071) genes, suggesting spurious exon usage calls occurring from expression-level differences. rMATS is competitive within the events it tests. rMATS tests orders of magnitude fewer number of events and genes, severely limiting its recall in the full universe. In the restricted universe, rMATS recall rises from 0.226 (full) to 0.425 (restricted), and the F1 reaches 0.523, the best of any tool, but only across the smaller universe of events it tested. GrASE focuses on precision in the precision-recall trade-off, yet it still recovers the most true events of any tool (11,146), with far fewer false positives than DEXSeq and an order of magnitude more detections than junction-based rMATS (Table 2).

*Table 2. Comparison of performance (adjusted p-value 0.01) across different software measured in units of exonic parts for GrASE and DEXSeq and units of events for rMATS.*

| Tool / Universe | Universe (tests) | Genes | TP | FP | FN | Precision | Recall | F1 |
|---|---|---|---|---|---|---|---|---|
| GrASE | 615,358 | 57,964 | 11,146 | 3,236 | 26,569 | 0.775 | 0.296 | 0.428 |
| GrASE (restricted) | 328,050 | 12,892 | 11,146 | 3,236 | 22,406 | 0.775 | 0.332 | 0.465 |
| DEXSeq | 615,358 | 57,964 | 10,812 | 9,306 | 26,903 | 0.537 | 0.287 | 0.374 |
| DEXSeq (restricted) | 393,423 | 14,678 | 10,812 | 9,306 | 26,018 | 0.537 | 0.294 | 0.380 |
| rMATS | 60,044 | 14,795 | 1,167 | 555 | 4,005 | 0.678 | 0.226 | 0.339 |
| rMATS (restricted) | 20,097 | 6,789 | 1,167 | 555 | 1,576 | 0.678 | 0.425 | 0.523 |

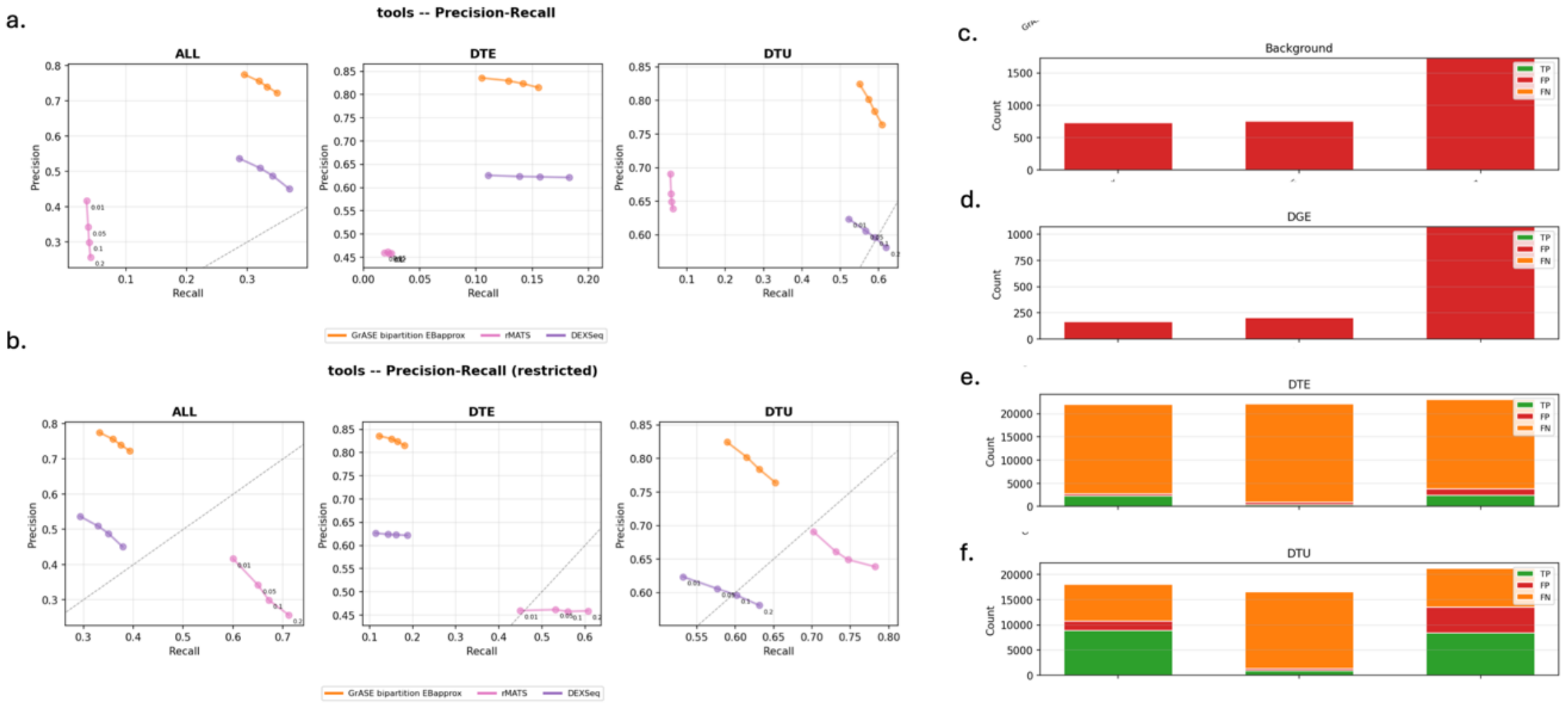

*Figure 7. Comparison of performance across different software measured in units of exonic parts for GrASE and DEXSeq and units of events for rMATS.*

The recall for DTE simulations is significantly lower than for DTU simulations for the exonic part tools (GrASE 0.102, DEXSeq 0.139, full universe). The reason for this is partially due to the penalizing way that we translate ground truths for exonic parts. The exonic part ground truth marks *every exonic part* of the changed transcript of DTEs as Positive, including constitutive and shared exons, resulting in a much larger Positive set that produces large FNs and lower recalls when any exonic part is missed. The restricted universe doesn't necessarily exclude these exonic parts, because due to the complex nested bubbles, almost all exons are part of some bubble, if not necessarily the Positive bubble that the transcript participates in, so many exons survive the restriction. For example, there are 21,574 exonic parts that are considered Positive as part of DTE transcripts, and restricting the universe to exonic parts tested by GrASE only reduces it by ~14% to 18,484.

In contrast, the junction ground truth only marks the junction-based events where the changed transcript participates in as Positive resulting in a small Positive set of 2,527 events. It also calls the whole event including exclusion and inclusion as one call by design, so regardless of which transcript is chosen as the changed transcript, skipping or including, it will be detected as a whole event. Thus, the DTE recall for rMATS is higher in the full universe, and rises further to 0.376 in the restricted universe.

The higher recall of rMATS on DTE genes can also be attributed to its higher power through the utilization of direct evidence of split reads across the junctions. To examine this effect, we directly compared the events that rMATS called TPs but GrASE and DEXseq missed. During this comparison between GrASE and rMATS, we discovered another side effect resulting from the different ways we translated ground truth for GrASE and DEXSeq compared to rMATS. If the DTE simulation chooses the skipped exon transcript to be the transcript with fold-change, we never marked the skipped exon as ground truth Positive for exonic part methods, as it is not part of the changed transcript. GrASE separates the tests between the two bipartitions ($D_1$ and $D_2$) and only one of those tests, the one that the changed transcript passes through, is considered Positive. In cases of skipped exons, often the distinct set for skipped exons is an empty set and cannot be tested, while the opposite distinct set for the included exons are non-empty and can be tested but will not have been marked Positive because it is not the transcript that was changed. So, if the skipped exon transcript is the changed transcript, the cassette exon itself will be marked negative while, all the remaining exon the transcript passes through will be marked positive. Among the 438 DTE true-positive events unique to rMATS that we could confidently map to exonic parts, 250 (57%) were such cases—the exonic parts that was called Positive in the rMATS ground truth, was considered Negative in the exonic part ground truth because it was not part of the transcript that was changed. Of those 250, DEXSeq incorrectly calls the cassette as significant in 133/250 (53%) cases, False Positives driven by its "other" rest-of-gene denominator shifting when one isoform changes. GrASE does so in only ~10% (24/250). For the rest of the cassette exons, the $lfc_{D_net}$ filter of GrASE removes 91 of them, because they have $lfc_{D_net}$ <= 0 (median -0.84).

The rest of the 188 loci are genuine misses where the mapped exon is considered Positive both by rMATS ground truth and in the exonic part ground truth as it is part of the transcript that was changed. Of those, GrASE detects the exon at 110 / 188 (59%) and misses 78 (41%). DEXSeq rescues 28 out of those 78 and also misses 50. So, 50 / 188 (27%) are missed by both GrASE and DEXSeq but caught by rMATS. These 78/188 or 50/188 show the power of direct evidence that junction split reads provide. Among the 78 GrASE-missed loci, 67 are at shared exonic parts where both the changed transcript and other unchanged transcripts cover the exonic part. The pooled read counts are diluted by the counts from unchanged transcripts, and we cannot tell the difference in the source of the reads. As a result, the pooled exon-level fold change barely shifts (median |log2fold| 0.31 vs 0.67 at detected loci). Whereas, with rMATS, one can distinguish the source of the reads by counting the reads that span alternative junctions directly, cutting through the diluting noise.

**The power of GrASE lies in the detection of alternative TSS or TTS events with high precision.**

Even though the limitation of exonic part methods was described clearly for DTE simulations in the earlier section, the overall metrics are higher for GrASE than for rMATS. This is due to the larger number of TPs detected by GrASE across both universes contributing to the overall metric. Most of these TPs are detected in alternative Transcription Start Site (TSS) or Transcription Termination Site (TTS) events. By design, junction-based methods such as rMATS cannot detect these events, because there are no incoming or outgoing junctions at the start or end of these events that these methods can rely on.

There are two kinds of bubbles in a classic splicing graph. Internal bubbles where both source and sink nodes are within the transcript body, are the classic alternative splicing events that junction-based methods are equipped to detect. TSS/TTS bubbles are bubbles where either the source is the abstract node R that represents the start of the transcript, or the sink is the abstract node L that represents the end of the transcript. These represent alternative TSS or TTS that are not modeled by junction-based methods and are largely under-detected in the literature. Although the two kinds of bubbles can be distinguished, there is no methodological difference in how each is processed within GrASE.

The coverage and performance of GrASE are dominated by TSS/TTS bubbles, with nine times more exonic parts in the TSS/TTS bubbles (14,846 DTU and 18,354 DTE positive exons in TSS/TTS vs. 1,729 DTU and 1,942 DTE positive exons in internal bubbles), and higher recall and F1s (Supplementary Table 2-3). On internal exons, GrASE has lower recall than on TSS/TTS bubbles, although the precision is higher. rMATS has both higher precision and higher recall than GrASE on internal events, in both DTE and DTU simulations. The reason rMATS achieves lower precision than GrASE on the whole internal events is due to the higher number of FPs rMATS accumulated in the DGE and Background simulations compared to GrASE (Supplementary Table 2).
On the TSS/TTS bubbles, GrASE achieves higher precision than DEXSeq, and rMATS cannot compete at all. Recall on DTE simulations is low for both GrASE and DEXSeq on both internal and TSS/TTS bubbles due to the diluted counts in shared exonic parts documented earlier (Supplementary Table 3).

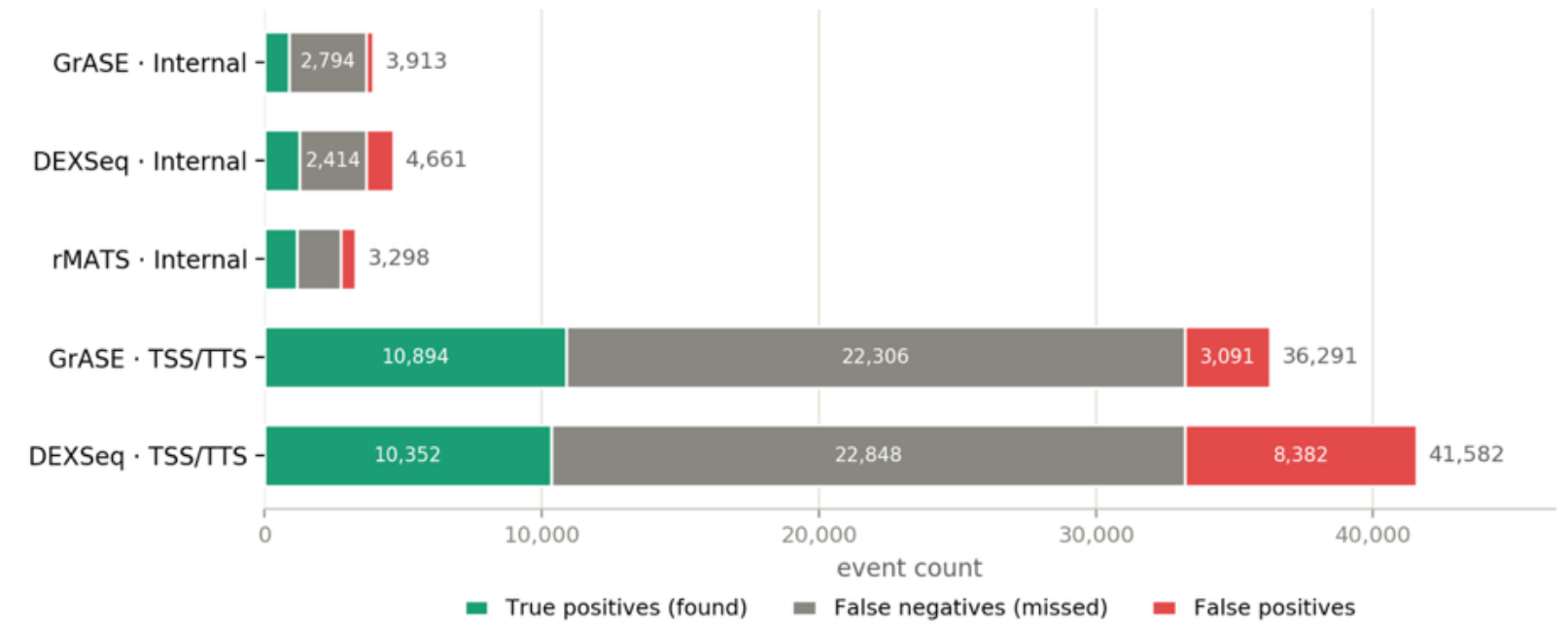

*Figure 8. Counts of prediction by ground truth by different software across internal and TSS/TTS bubbles.*

**Application to transcriptomes of human immune cell types.**

To show the utility of GrASE on detecting alternative splicing in empirical data, we applied the algorithm to the dataset from the Database of Immune Cell Expression, eQTLs, and Epigenomics (DICE) study (Schmiedel *et al.* 2018). DICE is a resource providing bulk RNA-seq profiles of sorted human immune cell populations from a cohort of healthy, genotyped donors. We analyzed five $CD4^{+}$ T cell populations from DICE, resting naive $CD4^{+}$ T cells (CD4), their activated counterpart (CD4_N_STIM), and three polarized effector lineages (TH1, TH2, and TH17). $CD4^{+}$ T cell activation and helper-lineage differentiation are well-established drivers of widespread transcriptional and isoform-level remodeling, making these populations a well-suited system for evaluating detection of differential splicing.

We structured the analysis around four pairwise contrasts. The first, CD4 vs CD4_N_STIM, isolates the transcriptional response to activation alone, before lineage commitment. The remaining three contrasts, CD4_N_STIM vs TH1, vs TH2, and vs TH17, each capture the changes accompanying differentiation into a specific effector lineage. Anchoring all three differentiation contrasts to the same activated precursor (CD4_N_STIM), controls for the shared activation signal and isolates the lineage-specific programs that distinguish TH1, TH2, and TH17 fates.

Significant path proportions were called with the condition that $lfc_{D_net} > 0$ and the shift in path proportion $\pi$ is greater than 0.1 (|delta_pi| >= 0.1) and at padj < 0.05, with nested BH and independent filtering. For the real-data analysis we report events at padj < 0.05, a slightly more permissive threshold than the padj < 0.01 used for the simulation calls to include condition on effect size.

*Table 3. Total number of exonic parts and significantly differentially used exonic parts in internal and TSS/TTS bubbles.*

| | Internal | | TSS/TTS | |
|---|---|---|---|---|
| **Contrast** | **total** | **significant (%)** | **total** | **significant (%)** |
| CD4 vs CD4_N_STIM | 28,281 | 934 (3.3%) | 416,059 | 15,642 (3.8%) |
| CD4_N_STIM vs TH1 | 28,036 | 989 (3.5%) | 414,346 | 15,842 (3.8%) |
| CD4_N_STIM vs TH2 | 28,172 | 1,035 (3.7%) | 415,209 | 15,573 (3.8%) |
| CD4_N_STIM vs TH17 | 28,060 | 984 (3.5%) | 414,160 | 15,614 (3.8%) |

Similar to the simulated data, we tested about 15 times more bipartition path proportions in TSS/TTS bubbles than in internal bubbles. The positive rate was similar between internal (3.3-3.7%) and TSS/TTS (3.8%) bubbles. The threshold on Path Proportion $\pi$ (|delta_pi| >= 0.1) reduced the positive rate from 15-19% to 3.3~3.8% (Table 3).

We confirmed the detection of several well-known positive controls. PTPRC (CD45, ENSG00000081237), the canonical alternative splicing gene in T cells, that undergoes well-documented isoform switching (CD45RA -> CD45RO) during T cell activation, was highly significant across all four contrasts. The two RNA-binding proteins CELF1 and CELF2, which are well-established regulators of alternative splicing in T cells were both significant in all four contrasts. CELF2 is known to directly regulate CD45 (PTPRC) alternative splicing during T cell activation. CD44 (ENSG00000026508), which is a classic marker of T cell activation and memory was significant across all four contrasts.

To distinguish differential exonic part usage from differential gene expression, we also ran DESeq2 on the same samples, calling significance at |log2FC| >= 1.5-fold, and padj < 0.05. We found that “DGE only” was the largest category with > 8,000 genes showing difference in gene expression but no detection of differential exon usage across all contrasts (Supplementary Table 4). About > 2,500 genes showed only differential exon usage and no differential gene expression, and about the same number showed both signatures. Genes detected in the DGE only category included cytokines, lineage master TFs, AP-1 complex, and NF-kB factors.

To focus on the functional relevance of differential exon usage, we ran an enrichment analysis excluding the genes detected with DGE, using GOseq (Young *et al.* 2010) to control for variation in the total number of exonic parts that arise from gene length and transcript complexity. “mRNA splicing, via spliceosome” was the single strongest term (FDR=0.00013), with genes such as DDX5, DDX46, DHX8/16/38, SNRNP200, SNRNP70, SNRPA/A1, SF1, SF3A1/2, PRPF3, SRSF4, TRA2A/B, RBM5, RBM22, SRRM1/2, and CRNKL1 (Figure 9). It was also the top hit in every individual contrast that had any signal. This is in contrast with the weak signal we found for this term when we included all the DGE genes (Supplementary Figure 4).

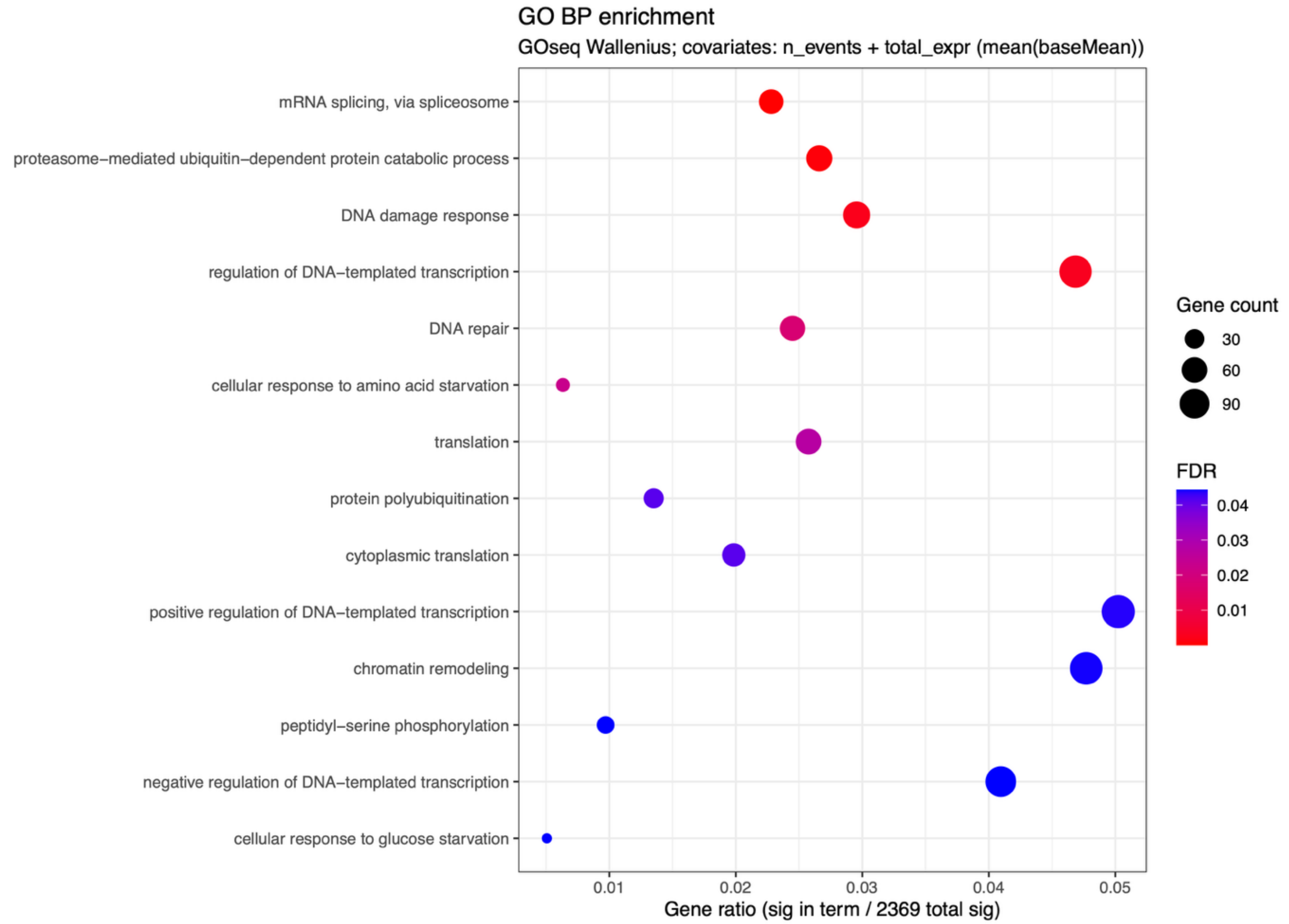

*Figure 9. GO terms Biological Processes enriched among the genes showing differential exon usage, but no differential gene expression pooled across the 4 contrasts.*

DDX5 and TRA2B carried very large effects across all four contrasts (|delta-pi| ~0.44), mainly driven by TSS/TTS change, without showing differential gene expression (Table 4). These are splicing factors with documented autoregulation and isoform control of apoptotic targets. For TRA2B, every significant event separated the productive multi-transcript path from a single retained-intron isoform, with at least one that could trigger Non-sense Mediated Decay (NMD). Across 37 genes in the "mRNA splicing, via spliceosome" term, 30 (~80%) shifted predominantly toward an unproductive minor isoform, suggesting a possible coordinated change in the productive-to-unproductive isoform ratio across the splicing machinery with T-cell state. (Supplementary Materials)

Other notable functions enriched were “DNA damage response”, with genes such as MCL1, CASP9 which are canonical apoptotic splice switches. “Nutrient-sensing / mTOR network” included MTOR, NPRL3, KICS2, FLCN, EIF2AK3 (PERK). Immune regulators were also detected such as RUNX1, with RUNX1a/b/c isoforms that well characterized in hematopoiesis, CIITA, with alternative promoters (pI/pIII/pIV) that are reported, and IKZF1 (Ikaros), whose dominant-negative splice isoforms (e.g. Ik6) are well known in lymphocyte development and leukemia.

*Table 4. Example genes with significantly differentially used path proportion. Estimated path proportion in different cell states.*

| Gene (event) | alternative isoform | rest CD4 | activated N_STIM | TH1 | TH2 | TH17 | direction |
|---|---|---|---|---|---|---|---|
| TRA2B pp1 | retained-intron / NMD (unproductive) | 0.47 | 0.16 | 0.44 | 0.52 | 0.47 | unproductive LOW in activated CD4. |
| DDX5 pp15 | retained-intron (unproductive) | 0.63 | 0.25 | 0.61 | 0.65 | 0.61 | unproductive LOW in activated CD4. |
| IKZF1 pp10 | alt-TSS truncated dominant-negative (productive) | 0.13 | 0.21 | 0.10 | 0.10 | 0.10 | dominant-negative HIGH in activated CD4 |

We looked at IKZF1 isoform switch in more detail to see if the isoform usage we detected is consistent with the known biology. The minor-path isoform (ENST00000492782) started at 50308748, downstream of the Matched Annotation from NCBI and EMBL-EBI (MANE) start 50304716 and had 4 CDS exons vs 7 in MANE, thus consistent with the N-terminally truncated, zinc-finger-deficient dominant-negative configuration. We found that the dominant-negative-type Ikaros isoform is highest in activated-but-undifferentiated CD4_N_STIM (pi 0.21) and lowest in the committed Th subsets (~0.10) (Table 4). A shift away from the dominant-negative toward full-length Ikaros as cells commit to a Th fate is consistent with the known role of IKZF1. This isoform level dynamic of IKZF1 across activated vs differentiated T-cell states we observed have not been reported before.

We also found some hits that are novel or understudied for their isoform usages in T-cell differentiation. Most noteworthy were the genes RC3H1 / RC3H2 (Roquin-1/2), which were significant across most contrasts (padj ~1e-28 to 1e-37). The two genes are central post-transcriptional repressors of ICOS/OX40 and T-cell tolerance, but their isoform functions are uncharacterized. We also detected METTL14, a m6A writer whose long/short 3'UTR isoforms with distinct functions were reported recently in hepatocellular carcinoma but have not been studied in T-cells.

# Conclusion

We present a unified computational framework, GrASE, based on set operations over exonic part paths along the splicing graph, for detecting alternative splicing (AS) in complex bubbles. Our methodological contribution is the efficient lower-set bipartitioning algorithm, which yields partitions with well-defined distinct exonic parts in time that we found is roughly linear in the number of alternative paths. This allows us to scale to genome-wide complex splicing events and increase the number and types of events that are tested. Another contribution is our observation and quantification of reference effect that can lead to false positives in any test built on compositional ratio, together with a heuristic correction based on the $lfc_{D_net}$ statistic that shows good control of this effect.
Our approach has clear limitations. We use a practical cap of maximum k=15 paths, which excludes about 3% of the larger bubbles. This means we cannot analyze the few most combinatorially complex loci, although this is a tunable trade-off between completeness and runtime. The recall on DTE events is especially low for GrASE, and this reflects a fundamental constraint. When a distinct exonic part is shared by both changed and unchanged transcripts, the pooled read count dilutes the signal and the source of individual reads cannot be recovered. Junction-based methods retain higher power here because split reads spanning alternative junctions carry direct evidence of which path produced them.

GrASE differs from existing software in two respects. First, rather than inferring the abundance of unobservable sub-transcript paths, it compares groups of paths through the shared and distinct exonic parts they contain, allowing direct quantification from mapped reads, and avoiding the estimation problems of deconvolving overlapping isoforms. Second, its lower-set bipartitioning provides a principled way to structure comparisons in complex bubbles where naïve one-versus-rest partitioning fails, while its reliance on exonic coverage rather than spanning junctions lets it test the large class of alternative TSS/TTS events that junction-based methods cannot detect at all. GrASE is therefore best viewed as complementary to junction-based approaches. The unique strength of GrASE lies in its significant expansion of testable space, especially the detection of alternative TSS and TTS events, that constitute a large fraction of alternative isoforms but have been understudied so far. Combining this larger testable space with its higher precision, GrASE detects more significant events accurately despite the trade-off of lower recall.

Anders S, Pyl PT, Huber W. HTSeq—a Python framework to work with high-throughput sequencing data. *Bioinformatics* 2015;**31**(2):166–9. https://doi.org/10.1093/bioinformatics/btu638.

Anders S, Reyes A, Huber W. Detecting differential usage of exons from RNA-seq data. *Genome Res* 2012;**22**(10):2008–17. https://doi.org/10.1101/gr.133744.111.

Aquino J, Witoslawski D, Park S *et al.* A novel splicing graph allows a direct comparison between exon-based and splice junction-based approaches to alternative splicing detection. *Brief Bioinform* (England) 2025;**26**(3). https://doi.org/10.1093/bib/bbaf204.

Bindreither D, Carlson M, Morgan M *et al. SplicingGraphs: Create, Manipulate, Visualize Splicing Graphs, and Assign RNA-Seq Reads to Them.*, version 1.52.0. R Package. n.d. https://doi.org/10.18129/B9.bioc.SplicingGraphs.

Brooks ME, Kristensen K, Benthem KJ van *et al.* glmmTMB Balances Speed and Flexibility Among Packages for Zero-inflated Generalized Linear Mixed Modeling. *R J* 2017;**9**(2):378–400. https://doi.org/10.32614/RJ-2017-066.

Davey BA, Priestley HA. *Introduction to Lattices and Order*. 2nd edn, Cambridge: Cambridge University Press, 2002. https://doi.org/10.1017/CBO9780511809088.

Heber S, Alekseyev M, Sze SH *et al.* Splicing graphs and EST assembly problem. *Bioinforma Oxf Engl* (England) 2002;**18 Suppl 1**:S181-188. https://doi.org/10.1093/bioinformatics/18.suppl_1.s181.

Katz Y, Wang ET, Airoldi EM *et al.* Analysis and design of RNA sequencing experiments for identifying isoform regulation. *Nat Methods* (United States) 2010;**7**(12):1009–15. https://doi.org/10.1038/nmeth.1528.

Love M, Soneson C, Patro R. Swimming downstream: statistical analysis of differential transcript usage following Salmon quantification [version 3; peer review: 3 approved]. *F1000Research* 2018;**7**(952). https://doi.org/10.12688/f1000research.15398.3.

Love MI, Huber W, Anders S. Moderated estimation of fold change and dispersion for RNA-seq data with DESeq2. *Genome Biol* 2014;**15**(12):550. https://doi.org/10.1186/s13059-014-0550-8.

Schmiedel BJ, Singh D, Madrigal A *et al.* Impact of Genetic Polymorphisms on Human Immune Cell Gene Expression. *Cell* 2018;**175**(6):1701-1715.e16. https://doi.org/10.1016/j.cell.2018.10.022.

Shen S, Park JW, Lu Z xiang *et al.* rMATS: Robust and flexible detection of differential alternative splicing from replicate RNA-Seq data. *Proc Natl Acad Sci* 2014;**111**(51):E5593–601. https://doi.org/10.1073/pnas.1419161111.

Simes RJ. An improved Bonferroni procedure for multiple tests of significance. *Biometrika* 1986;**73**(3):751–4. https://doi.org/10.1093/biomet/73.3.751.

Smyth GK. Linear Models and Empirical Bayes Methods for Assessing Differential Expression in Microarray Experiments. *Stat Appl Genet Mol Biol* 2004;**3**(1). https://doi.org/10.2202/1544-6115.1027.

Sterne-Weiler T, Weatheritt RJ, Best AJ *et al.* Efficient and Accurate Quantitative Profiling of Alternative Splicing Patterns of Any Complexity on a Laptop. *Mol Cell* 2018;**72**(1):187-200.e6. https://doi.org/10.1016/j.molcel.2018.08.018.

Trincado JL, Entizne JC, Hysenaj G *et al.* SUPPA2: fast, accurate, and uncertainty-aware differential splicing analysis across multiple conditions. *Genome Biol* 2018;**19**(1):40. https://doi.org/10.1186/s13059-018-1417-1.

Van den Berge K, Soneson C, Robinson MD *et al.* stageR: a general stage-wise method for controlling the gene-level false discovery rate in differential expression and differential transcript usage. *Genome Biol* 2017;**18**(1):151. https://doi.org/10.1186/s13059-017-1277-0.

Vaquero-Garcia J, Barrera A, Gazzara MR *et al.* A new view of transcriptome complexity and regulation through the lens of local splicing variations. *eLife* 2016;**5**:e11752. https://doi.org/10.7554/eLife.11752.

Wang Q, Rio DC. JUM is a computational method for comprehensive annotation-free analysis of alternative pre-mRNA splicing patterns. *Proc Natl Acad Sci* 2018;**115**(35):E8181–90. https://doi.org/10.1073/pnas.1806018115.

Xing Y, Resch A, Lee C. The Multiassembly Problem: Reconstructing Multiple Transcript Isoforms From EST Fragment Mixtures. *Genome Res* 2004;**14**(3):426–41. https://doi.org/10.1101/gr.1304504.

Young MD, Wakefield MJ, Smyth GK *et al.* Gene ontology analysis for RNA-seq: accounting for selection bias. *Genome Biol* 2010;**11**(2):R14. https://doi.org/10.1186/gb-2010-11-2-r14.